\documentclass[notitlepage,letterpaper, 11pt]{article}
\usepackage[ansinew]{inputenc} 
\usepackage{amsmath}
\usepackage{amsfonts}
\usepackage{amssymb}
\usepackage{graphicx}
\usepackage{subfig}
\usepackage{cancel}
\usepackage{booktabs}
\usepackage{color}
\usepackage{bigints}
\usepackage[top=1cm, bottom=2cm,outer=2cm, inner=2cm]{geometry}
\usepackage{amsthm}
\usepackage{lineno}

\begin{document}

\title {Karmarkar scalar condition}
\author{
J. Ospino\thanks{j.ospino@usal.es} \\
\textit{Departamento de Matem\'atica Aplicada and} \\
\textit{Instituto Universitario de F\'isica Fundamental y Matem\'aticas,}  \\ 
\textit{ Universidad de Salamanca, Salamanca-Spain;} \\
 L.A. N\'u\~nez\thanks{lnunez@uis.edu.co} \\
\textit{Escuela de F\'isica, Universidad Industrial de Santander, } \\ \textit{ Bucaramanga-Colombia and} \\ \textit{Departamento de F\'isica, Universidad de Los Andes,} \\ \textit{ M\'erida-Venezuela.}
}
\maketitle

\begin{abstract}
In this work we present the Karmarkar condition in terms of the structure scalars obtained from the orthogonal decomposition of the Riemann tensor. This the new expression becomes an algebraic relation among the physical variables, and not a differential equation between the metric coefficients. By using the Karmarkar scalar condition we implement a method to obtain all possible embedding class I static spherical solutions, provided the energy density profile is given. We also analyse the dynamic adiabatic case and show the incompatibility of the Kamarkar condition with several commonly assumed simplifications to the study of gravitational collapse. Finally, we consider the dissipative dynamic Karmarkar collapse and find a new solution family.
\end{abstract}
\maketitle

\section{Introduction}
General Relativity is living unprecedented times, witnessing the transformation of exotic objects --such as black holes-- and feeble phenomena --like gravitational waves-- from mathematical curiosities to observable physical entities.
 
There are many notable attempts to explore the properties of physically viable solutions (numeric \& analytic) describing either static, stationary, or collapsing relativistic compact objects. All known exact solutions have been obtained by imposing some restrictions, such as symmetry conditions on the metric, the algebraic structure of the Riemann tensor, new coupled field equations, meaningful equations of state for the matter variables, or selecting particular initial and boundary conditions, to mention the most common strategies. 

Einstein's covariant mathematical description of gravitation contrasts with solution obtained which are strongly dependent on the coordinate basis. It is not always easy to understand the qualitative features that these coordinate-prone solutions might possess, and the analysis of their general properties could reveal unforeseen features of the theory. Thus it is useful to study the general properties through a coordinate independent formalism. 

We have recently implemented a tetrad formalism by an orthogonal splitting of the Riemann tensor. We introduced a full set of equations equivalent to the Einstein system and applied it to the spherical case, showing that it is possible to obtain relevant information from self-gravitating systems \cite{OspinoHernandezNunez2017, OspinoEtal2018}. This formalism provides coordinate-free results expressed in terms of structure scalars closely related to the kinematical and physical properties of the fluid. 

In this short paper, we shall explore the consequences of imposing the well-known Karmarkar condition \cite{karmarkar1948}, which implies that a curved four-dimensional metric can be embedded into a five-dimensional pseudo-Euclidean space-time.  The Karmarkar condition provides a geometric relation between the metric functions and their derivatives; thus, one can choose one of the metric functions and generate the other. In the case of an isotropic static fluid sphere --Pascalian matter distribution--  the Karmarkar condition leads to either a Schwarzschild --homogeneous conformally-flat bounded solution--  or a Kohler-Chao solution --a non-conformally flat unbounded solution \cite{GuptaGupta1986}--. However, for static anisotropic matter configurations, it provides a geometrical mechanism for implementing equations of state relating the radial and the tangential pressures.   

\begin{figure}
    \centering
    \includegraphics[scale=0.28]{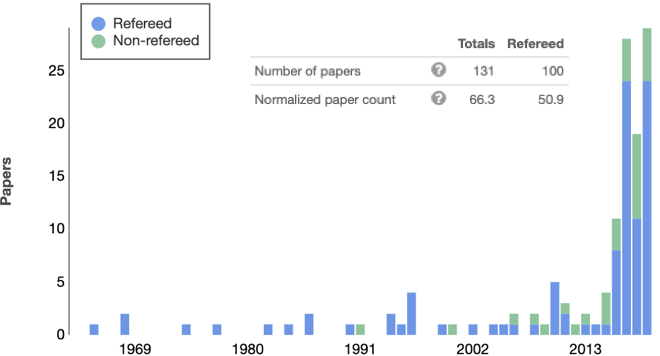}
    \includegraphics[scale=0.28]{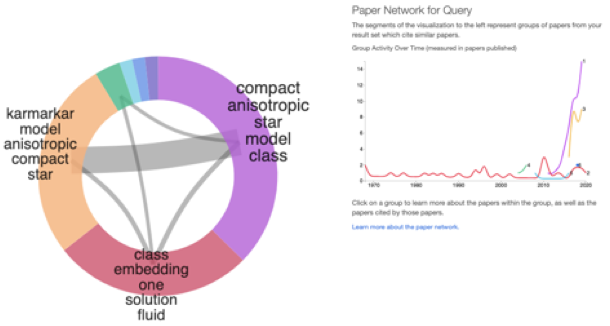}
    \caption{A search in the astrophysics data system (\texttt{https://ui.adsabs.harvard.edu/}) for ``Karmarkar Condition'' OR ``embedding class one'' OR ``embedding class 1'' from 1948 to 2019 gives 131 publications (left plate). Most of them are recent and devoted to describing compact objects with anisotropic equations of state. On the right plate, the publications are grouped, attending to the commonly cited references in each paper}
    \label{fig:KarmarkarNet}
\end{figure}

As pointed out by B.V. Ivanov \cite{Ivanov2017}, the Karmarkar initial embedding motivation changes into a geometical method that generates matter configurations that may represent compact astrophysical objects.  As shown in figure \ref{fig:KarmarkarNet} Karmarkar's condition has experimented a recent boom, with more than 70 publications in the last three years, most of them devoted in describing anisotropic compact objects. There are many interesting models of possible compact objects, depending on the variety of the metric function selected as input: rational functions\cite{MauryaEtal2016B, SinghPant2016B, SinghEtal2017, Bhar2017, BharSinghManna2017, BharEtal2017, SinghPantGovender2017}, polynomials \cite{SinghBharPant2016, BharEtal2016, SinghPradhanPant2017, SinghPant2016, MauryaEtal2016C}, trigonometric \cite{SinghPantGovender2017B, SinghPantTroconis2017, Fuloria2017, SinghMuradPant2017} and hyperbolic functions \cite{MauryaRatanpalGovender2017, MauryaMaharaj2017, MauryaEtal2017}. Recently, there have been some explorations of the consequences of the Karmarkar conditions on stellar structure models in modified theories of gravity \cite{DebEtal2019, AbbasNazar2019}.  

In the next section, we briefly describe the scalar formalism we use. Following, in section \ref{SectKarmarkarCond}, we study the static and dynamic (adiabatic and dissipative scenarios) Karmarkar solutions. For the static case, we implement an algorithm to generate any Karmarkar spherical static anisotropic solution given the energy density profile. Regarding the dynamic adiabatic assumption, we show how restrictive the Karmarkar condition may be, and for the corresponding dissipative environment, we found a new family of dynamical radiating Karmarkar line-elements.  Finally, our last section wraps-up some remarks and conclusions.

\section{The structure scalar strategy and the general formalism}
As we mentioned above, the strategy we shall follow is to formulate two independent sets of equations, --expressed in terms of scalar functions--, which contain the same information as the Einstein system.

Let us choose an orthogonal unitary tetrad:
\begin{equation}
\label{TetradGen}
e^{(0)}_\alpha~=~V_\alpha, \, e^{(1)}_\alpha~=~K_\alpha, \, e^{(2)}_\alpha~=~L_\alpha \; \; \mathrm{and}  \; e^{(3)}_\alpha~=~S_\alpha.
\end{equation}

As usual, $\eta_{(a)(b)}~=~ g_{\alpha\beta} e_{(a)}^\alpha e_{(b)}^\beta$, with $a=0,\,1,\,2,\,3$, i.e. latin indices label different vectors of the tetrad. Thus, the tetrad satisfies the standard relations:
\begin{eqnarray}
V_{\alpha}V^{\alpha} &=& -K_{\alpha}K^{\alpha} = -L_{\alpha}L^{\alpha} = -S_{\alpha}S^{\alpha} = -1\,, \nonumber \\
V_{\alpha}K^{\alpha} &=& V_{\alpha}L^{\alpha} = V_{\alpha}S^{\alpha} = K_{\alpha}L^{\alpha} = K_{\alpha}S^{\alpha} = S_{\alpha}L^{\alpha} = 0\,.  \nonumber
\end{eqnarray} 
With the above tetrad (\ref{TetradGen}) we shall also define the corresponding directional derivative operators
\begin{equation}
\label{DirectionalDerivatives}
f^{\bullet} = V^{\alpha} \partial_{\alpha}f ; \quad f^{\dag} = K^{\alpha} \partial_{\alpha}f  \quad \mathrm{and} \quad f^{\ast} = L^{\alpha}\partial_{\alpha}f .
\end{equation}

The first set can be considered  purely geometrical and emerges from the projection of the Riemann tensor along the tetrad \cite{Wald2010}, i.e.
\begin{equation}
    \label{RiemannProj}
2V_{\alpha\, ;[\beta ;\gamma]} = R_{\delta \alpha \beta \gamma} V^{\delta}, 
\quad   2K_{\alpha\, ;[\beta ;\gamma]} = R_{\delta \alpha \beta \gamma} K^{\delta}, \quad 
2L_{\alpha \, ;[\beta ;\gamma]} = R_{\delta \alpha \beta \gamma} L^{\delta}   \quad \mathrm{and} \quad   
2S_{\alpha\, ;[\beta ;\gamma]} = R_{\delta \alpha \beta \gamma} S^{\delta}\,  ; 
\end{equation}
where $e^{(a)}_{\alpha \, ;\beta \gamma}$ are the second covariant derivatives of each tetrad (\ref{tetrad}) vector indicated with $a = 0,1,2,3.$

The second set emerges from the Bianchi identities:
\begin{equation}
\label{BianchiIdent}
R_{\alpha \beta [\gamma \delta \, ;  \mu]} =
R_{\alpha \beta \gamma \delta \, ;  \mu}  + R_{\alpha \beta  \mu \gamma  \, ; \delta} + R_{\alpha \beta \delta \mu \, ; \gamma } = 0 \, .
\end{equation}

\subsection{The tetrad, the source and the kinematical variables}
To proceed with the above objective we shall restrict to a spherically symmetric line element given by
\begin{equation}
\mathrm{d}s^2=-A(r,t)^2\mathrm{d}t^2 +B(r,t)^2 \mathrm{d}r^2 +R(r,t)^2 (\mathrm{d}\theta ^2+\sin^2(\theta)\mathrm{d}\phi ^2)\, ,
\label{SphericMetric}
\end{equation}
where the coordinates are: $x^0=t$, $x^1=r$, $x^2=\theta$.

In this case the tetrad is:
\begin{equation}
\label{tetrad}
V_{\alpha} = \left(-A,0,0,0\right),\,
K_{\alpha}=\left(0,B,0,0\right), \,
L_{\alpha}=\left(0,0,R,0\right), \, S_{\alpha} =
\left(0,0,0,R \sin(\theta)\right) \,,
\end{equation}
and their covariant derivatives can be written as:
\begin{eqnarray}
\label{CovDTetrad}
V_{\alpha;\beta}&=&-a_1K_\alpha V_\beta+\sigma_1K_\alpha K_\beta+\sigma_2(L_\alpha L_\beta+S_\alpha S_\beta),  \nonumber \\
K_{\alpha;\beta}&=&-a_1V_\alpha V_\beta+\sigma_1 V_\alpha K_\beta+J_1(L_\alpha L_\beta+S_\alpha S_\beta),  \\
L_{\alpha;\beta} & = &\sigma_2 V_\alpha L_\beta-J_1K_\alpha L_\beta+J_2S_\alpha S_\beta
\quad \mathrm{and} \quad S_{\alpha;\beta} = \sigma_2 V_\alpha S_\beta-J_1K_\alpha S_\beta -J_2L_\alpha S_\beta \,. \nonumber
\end{eqnarray}
Where: $J_1$, $J_2$, $\sigma_{1}$,  $\sigma_{2}$ and $a_1$ are expressed in terms of the metric functions and their derivatives as:
\begin{equation}
    \label{metricquantities}
J_1=\frac{1}{B}\frac{R^{\prime}}{R} \,, \quad 
J_2 = \frac{1}{R}\cot(\theta)\,, \quad
\sigma_{1} = \frac{1}{A}\frac{\dot{B}}{B}\,, \quad
\sigma_{2} = \frac{1}{A}\frac{\dot{R}}{R} \quad \mathrm{and} \quad  a_1=\frac{1}{B}\frac{A^{\prime}}{A} \,, 
\end{equation}
with primes and dots representing respectively, radial and time derivatives.

As we mentioned before we shall take as our source a bounded, spherically symmetric, locally anisotropic, dissipative, collapsing matter configuration, described by a general energy momentum tensor, written in the ``canonical'' form, as:
\begin{equation}
{T}_{\alpha\beta}= (\rho+P) V_\alpha V_\beta+P g _{\alpha \beta} +\Pi_{\alpha \beta}+\mathcal{F}_\alpha V_\beta+\mathcal{F}_\beta V_\alpha .
\label{EnergyTensor}
\end{equation}

It is immediately seen that the physical variables can be defined --in the Eckart frame  where fluid elements are at rest-- as:
\begin{equation}
\rho = T_{\alpha \beta} V^\alpha V^\beta, \quad 
\mathcal{F}_\alpha=-\rho V_\alpha - T_{\alpha\beta}V^\beta, \quad 
P = \frac{1}{3} h^{\alpha \beta} T_{\alpha \beta} \quad \mathrm{and} \quad \Pi_{\alpha \beta} = h_\alpha^\mu h_\beta^\nu \left(T_{\mu\nu} - P h_{\mu\nu}\right)\,,
\end{equation}
with $h_{\mu \nu}=g_{\mu\nu}+V_\nu V_\mu$.

As can be seen from the condition $\mathcal{F}^{\mu} V_{\mu}=0$, and the symmetry of the problem, Einstein's equations imply $T_{03}=0$, thus:
\begin{equation}
\mathcal{F}_{\mu}=\mathcal{F}K_{\mu} \quad \Leftrightarrow \quad \mathcal{F}_{\mu}= \left(0, \frac{\mathcal{F}}{B}, 0, 0\right) .
\label{energyflux}
\end{equation}
Clearly $\rho$ is the energy density (the eigenvalue of $T_{\alpha\beta}$ for eigenvector $V^\alpha$), $\mathcal{F}_\alpha$ represents the  energy flux four vector;  $P$ corresponds to the isotropic pressure, and $\Pi_{\alpha \beta}$ is the anisotropic tensor, which can be  expressed as
\begin{equation}
\Pi_{\alpha \beta}= \Pi_{1}\left(K_\alpha K_\beta -\frac{h_{\alpha \beta}}{3}\right)
\label{anisotropictensor},
\end{equation}
with
\begin{equation}
\Pi_{1}=\left(2K^{\alpha} K^{\beta} +L^{\alpha} L^{\beta}\right)  T_{\alpha \beta}.
\label{Pi1}
\end{equation}
Finally, we shall express the kinematical variables (the four-acceleration, the expansion scalar and the shear tensor) for a self-gravitating fluid as:
\begin{eqnarray}
a_\alpha&=&V^\beta V_{\alpha;\beta}=a K_\alpha =\left(0, \frac {A^{\prime} }{A },0,0\right),
\label{aceleration} \\
\Theta&=&V^\alpha_{;\alpha} =\frac{1}{A}\,\left(\frac{\dot{B}}{B}+\frac{2\dot{R}}{R}\right), \label{theta} \\
\sigma&=&\frac{1}{A}\left(\frac{\dot{B}}{B} -\frac{\dot{R}}{R}\right)\,.  \label{shear}
\end{eqnarray}

\subsection{The splitting of the Riemann tensor and structure scalars}

In this section we shall introduce a set of scalar functions --the structure scalars--  obtained from the orthogonal splitting of the Riemann tensor (see \cite{GarciaParrado2007,HerreraEtal2009B,HerreraEtal2014}) which has proven to be very useful in expressing the Einstein Equations.

Following \cite{GarciaParrado2007}, we can express the splitting of the Riemann tensor as:
\begin{eqnarray}
R_{\alpha \beta \mu \nu}&=&2V_\mu V_{[\alpha}Y_{\beta] \, \nu}+2h_{\alpha[\nu}X_{\mu] \,  \beta}+2V_\nu V_{[\beta}Y_{\alpha] \, \mu}
+ h_{\beta\nu}(X_0 \, h_{\alpha\mu}-X_{\alpha\mu})+h_{\beta\mu}(X_{\alpha\nu} -X_0 \, h_{\alpha\nu}) \nonumber \\
& & \qquad + 2V_{[\nu} Z_{ \, \mu]}^{\gamma}\varepsilon_{{\alpha \beta \gamma}} +2V_{[\beta} Z_{{\,  \ \alpha]}}^{{\gamma }}\ \varepsilon_{{\mu \nu \gamma}} \,,
\end{eqnarray}
with $\varepsilon_{\mu \nu \gamma} = \eta_{\phi \mu \nu \gamma} V^{\phi}$, and  $ \eta_{\phi \mu \nu \gamma}$ the Levi-Civita 4-tensor. The corresponding Ricci contraction for the above Riemann tensor can also be written as:
\begin{equation}
R_{\alpha\mu} =  Y_0 \, V_\alpha V_\mu-X_{\alpha \mu}-Y_{\alpha\mu} +X_0 \, h_{\alpha\mu}  +Z^{\nu \beta} \varepsilon_{\mu \nu \beta}V_{\alpha} 
+V_{\mu} Z^{\nu \beta} \varepsilon_{\alpha \nu \beta} \, ;
\end{equation}
where the quantities: $Y_{\alpha\beta}$, $X_{\alpha\beta}$ and $Z_{\alpha\beta}$ can be expressed as
\begin{eqnarray}
Y_{\alpha\beta}&=&\frac{1}{3}Y_0 \, h_{\alpha\beta} +Y_1\left[K_\alpha K_\beta-\frac{1}{3} h_{\alpha\beta}\right], \quad
X_{\alpha\beta} = \frac{1}{3} X_0 \, h_{\alpha\beta} +X_1\left[K_\alpha K_\beta-\frac{1}{3} h_{\alpha\beta}\right] \quad \mathrm{and}  \\
Z_{\alpha\beta}&=&Z \, (L_\alpha S_\beta-L_\beta S_\alpha)\,,
\end{eqnarray}
with
\begin{equation}
Y_0 = 4\pi(\rho+3P), \quad Y_1=\mathcal{E}_1-4\pi \Pi_1, \quad
X_0=8\pi \rho\,, X_1 = -(\mathcal{E}_1+4\pi \Pi_1) \quad
\mathrm{and} \quad  Z= 4 \pi \mathcal{F} \,,\label{varfis}
\end{equation}
and the electric part of the Weyl tensor is written as
\begin{equation}
E_{\alpha\beta}=C_{\alpha\nu\beta\delta}V^\nu V^\delta = \mathcal{E}_1\left[K_\alpha K_\beta-\frac{1}{3} h_{\alpha\beta}\right]\,.
\end{equation}

\subsection{Projections of Riemann tensor}
From the above system (\ref{RiemannProj}), by using the covariant derivative of equations (\ref{CovDTetrad}) and the projections of the orthogonal splitting of the Riemann tensor, we obtain the  first set of independent equations, for the spherical case, in terms of  $J_1$, $J_2$, $\sigma_{1}$,  $\sigma_{2}$, and $a_1$, (defined in (\ref{metricquantities})) and their directional derivatives, i.e.
\begin{eqnarray}
\sigma^{\bullet}_{1} -a_1^\dag-a_1^2+\sigma_1^2&=&-\frac{1}{3}(Y_0+2Y_1) \, , \label{ecR1} \\
\sigma^{\bullet}_{2} +\sigma_2^2-a_1J_1&=&\frac{1}{3}(Y_1-Y_0) \, , \label{ecR2}  \\
\sigma_2^\dag+J_1(\sigma_2-\sigma_1)&=&-Z \, , \label{ecR3} \\
J^{\bullet}_{1} +J_1\sigma_2-a_1\sigma_2&=&-Z \, , \label{ecR4} \\
J_1^\dag+J_1^2-\sigma_1 \sigma_2&=&\frac{1}{3}(X_1-X_0) \, , \label{ecR5} \\
J^{\bullet}_{2} +J_2\sigma_2&=&0 \, , \label{ecR6} \\
J_2^\dag+J_1J_2&=&0 \qquad \mathrm{and}  \label{ecR7} \\
J_1^2-\frac{1}{R^2}-\sigma_2^2&=&-\frac{1}{3}(X_0+2X_1) \,. \label{ecR8}
\end{eqnarray}

\subsection{Equations from Bianchi identities}
The second set of equations for the spherical case, emerge from the independent Bianchi identities (\ref{BianchiIdent}), and can be written as:
\begin{eqnarray}
a_1[-X_0+X_1&-&Y_0+Y_1]+3J_1Y_1+3Z^{\bullet} \nonumber \\
\qquad \qquad +6Z \sigma_1&+&3Z\sigma_2-Y_0^\dag+Y_1^\dag=0 \,,
\label{Bianchi1} \\
X^{\bullet}_0 -X^{\bullet}_1 -6a_1Z &-&3J_1Z  +\left[Y_0 -Y_1 -X_1\right] \sigma_1  
\nonumber \\
 +\left[Y_0 +2Y_1 -X_1\right] \sigma_2 &+&X_0[\sigma_1+\sigma_2]-3Z^\dag=0 \,, \label{Bianchi2} \\
X^{\bullet}_0 +2 X^{\bullet}_1 &+&2X_0\sigma_2 -6J_1Z \nonumber \\
+[4X_1 &+& 2Y_0 - 2Y_1] \sigma_2=0, \quad \mathrm{and}   \label{Bianchi3} \\  
X_0^\dag+2X_1^\dag&+&6J_1X_1+6Z\sigma_2=0 \,. \label{Bianchi4}
\end{eqnarray}

\section{Kamarkar condition}
\label{SectKarmarkarCond}
As it is well-known, a four-dimensional curved space-time can be embedded in a five-dimensional pseudo-Euclidean space whenever it satisfies the Karmarkar condition which can be stated as\cite{karmarkar1948}:
\begin{equation}\label{KamarkarC}
  R_{0303}R_{1212} -R_{0101}R_{2323} -R_{0313}R_{0212} = 0 \, ,
\end{equation}
and provides a geometrical mechanism to implement equations of state relating the radial and the tangential pressures. 

\subsection{Differential Karmarkar conditions}
When considering a line element (\ref{SphericMetric}), Karmarkar's condition (\ref{KamarkarC}) leads to 
\begin{eqnarray}
\label{KarmarkarCondSchwComovDyna}
&& \left[B^2 ( \dot{R}^2 +A^2)) A A^{\prime \prime} -(R^{\prime})^2 A^2 
+((R^{\prime})^2 A^2  -B^2 ( \dot{R}^2 +A^2)\right] B \ddot{B}
+ \left[ \dot{R} B^2 \dot{A} -\ddot{R} B^2 A  +R^{\prime} A^{\prime} A^2 \right] A R^{\prime \prime} \nonumber \\
&& +B^2 A^2 (\dot{R}^{\prime})^2
-(2 \dot{B} R^{\prime} A  +2  \dot{R} B A^{\prime}) A B \dot{R}^{\prime}
+ (R^{\prime} B^{\prime} A^2+ \dot{R} B^2 \dot{B}) B \ddot{R}
+(A \dot{B} -B \dot{A}) A \dot{B}  (R^{\prime})^2 \nonumber \\
&& + \dot{R} A B (A^{\prime} \dot{B} -\dot{A} B^{\prime}) R^{\prime}
-((B^{\prime} A A^{\prime} -B (A^{\prime})^2)  \dot{R}^2 +A (B^{\prime} A^{\prime} A^2 -B^2 \dot{B} \dot{A})) B  = 0 \, , 
\end{eqnarray}
as shown in reference \cite{NaiduGovenderMaharaj2018}, for the particular case of $A(r,t) = \tilde{A}(r)$,  $B(r,t) = \tilde{B}(r)f(t)$ and $R(r,t)~=~r \tilde{B}(r)f(t)$. 

If we examine a much simpler metric like
\begin{equation}
\mathrm{d}s^2=-\mathrm{e}^{\nu(r,t)}\mathrm{d}t^2 +\mathrm{e}^{\lambda(r,t)} \mathrm{d}r^2 +r^2 (\mathrm{d}\theta ^2+\sin^2(\theta)\mathrm{d}\phi ^2)\, ,
\label{SchwSphericMetric}
\end{equation}
the Karmarkar condition (\ref{KamarkarC}) can be written as
\begin{equation}
2\nu^{\prime \prime}\mathrm{e}^{\nu} -\left((\nu^{\prime})^2 + 2\nu^{\prime \prime}\right)\mathrm{e}^{(\nu -\lambda)}
-\nu^{\prime}\mathrm{e}^{\nu} \left( \lambda^{\prime}-\nu^{\prime} \right)
-\left(\mathrm{e}^{\lambda} -2\right)\dot{\lambda}^2
+\dot{\nu}\dot{\lambda}\left(\mathrm{e}^{\lambda} -1\right)
-2\ddot{\lambda}\left( \mathrm{e}^{\lambda} - 1\right) = 0 \, .
\label{KarmarkarCondSchwCurvDyna}    
\end{equation}
Which, in the static case, leads to the differential equation
\begin{equation}
\frac{2\nu^{\prime \prime}}{\nu^{\prime}} +\nu^{\prime} = \frac{\lambda^{\prime} \mathrm{e}^{\lambda} }{\mathrm{e}^{\lambda} -1} \, ,
\label{DiffKarmarkarStatic}    
\end{equation}
the most common expression for the Karmarkar condition examined in the literature (see references \cite{MauryaEtal2016B} through \cite{MauryaEtal2017}). 

If we provide a particular $\lambda$-function -- listed in table \ref{TableMetricsX0}-- we can obtain the other metric coefficient $\nu$ and then investigate the type of material described by this line-element. Thus, again, the Karmarkar condition implements a geometrical method to generate anisotropic equations of state, and has boomed a profusion of possible realistic models for compact objects. Unfortunately, (\ref{KamarkarC}), the models generated are coordinate dependent, and the general properties obtained are heavily conditioned from this fact.

\subsection{Scalar Karmarkar conditions}
This coordinate dependence can be overcome in the tetrad framework by projecting the Riemman tensor as 
\begin{eqnarray}
      &&   R_{\alpha\beta\mu\nu}V^\alpha V^\mu S^\beta S^\nu R_{\gamma\delta\sigma\rho} K^\gamma K^\sigma L^\delta L^\rho - R_{\alpha\beta\mu\nu}V^\alpha V^\mu K^\beta K^\nu R_{\gamma\delta\sigma\rho} L^\gamma L^\sigma S^\delta S^\rho  \nonumber \\
          & & \qquad \qquad -R_{\alpha\beta\mu\nu}V^\alpha K^\mu S^\beta S^\nu R_{\gamma\delta\sigma\rho} V^\gamma L^\sigma K^\delta L^\rho = 0
\end{eqnarray}
and, from equations (\ref{varfis}) assuming spherical symmetry, it can be reduced to a simple algebraic scalar relation among several physical variables:
\begin{equation}
Y_0 X_1+(X_0+X_1)Y_1=-3Z^2 \, .
\label{KarmaKcondTetrad}
\end{equation}
Notice that this scalar relation among the physical variables defined in equations (\ref{varfis}), despite its simplicity, is valid for any dynamic and dissipative spherical matter distribution described by (\ref{SphericMetric}). In the next sections we shall use (\ref{KarmaKcondTetrad}) to study, both the static and dynamic (adiabatic and dissipative) cases. 
\subsection{The static case}
Employing the above-sketeched scalar formalism and assuming the condition (\ref{KarmaKcondTetrad}), we shall find the most general static, spherically symmetric anisotropic Karmarkar solution.  

For the line element  (\ref{SphericMetric}) we can assume, without any loss of generality, $R=r$ and integrate (\ref{ecR2}) to obtain:
\begin{equation}
A=C_1 e^{\int \frac{B^2 r}{3}(Y_0-Y_1)dr} \,,
\label{Aest1}
\end{equation}
where $C_1$ is a constant of integration.  Next, from equation (\ref{ecR4}) it follows at once that:
\begin{equation}
B^2=\frac{1}{1-\frac{r^2}{3}(X_0+2X_1)}\label{BestB1}\,.
\end{equation}
Clearly, these metric elements (\ref{Aest1}) and (\ref{BestB1}) --expressed in terms of the structure scalars $X_1$  and $Y_0-Y_1$--, describe any static anisotropic matter distribution \cite{HerreraOspinoDiPrisco2008}.

The equivalent Einstein system of equations (\ref{ecR1})-(\ref{Bianchi4}) can be written, for the static case, as:
\begin{eqnarray}
a_1J_1&=&\frac{1}{3}(Y_0-Y_1) \, , \label{ecR2s}  \\
J_1^\dag+J_1^2&=&\frac{1}{3}(X_1-X_0) \, , \label{ecR5s} \\
J_1^2-\frac{1}{R^2}&=&-\frac{1}{3}(X_0+2X_1) \, , \label{ecR8s}\\
a_1[-X_0+X_1&-&Y_0+Y_1]+3J_1Y_1=(Y_0-Y_1)^\dag \, ,
\label{Bianchi1s} \\
X_0^\dag+2X_1^\dag&+&6J_1X_1=0 \,, \label{Bianchi4s}
\end{eqnarray}
and the Karmarkar condition (\ref{KarmaKcondTetrad}) takes the form of
\begin{equation}
(Y_0-Y_1)X_1+(X_0+2X_1)Y_1=0 \, .\label{Kcons}
\end{equation}

Now, integrating equation (\ref{Bianchi4}) we find
\begin{equation}
X_1=\frac{3}{2r^3}\int X_0 r^2 dr-\frac{1}{2}X_0 \,.
\label{X1es} 
\end{equation}
On the other hand, by using equations (\ref{ecR2s})-(\ref{ecR8s}) together with (\ref{Bianchi4s})-(\ref{Kcons}), equation (\ref{Bianchi1s}) can be written as
\begin{equation}
(Y_0-Y_1)^\dag=\left(\frac{J_1^\dag}{J_1}+J_1+\frac{1}{2}\frac{(X_0+2X_1)^\dag}{X_0+2X_1}\right)(Y_0-Y_1)-\frac{1}{3J_1}(Y_0-Y_1)^2 \, .
\label{Y0Y1}
\end{equation}
Now, integrating (\ref{Y0Y1})
\begin{equation}
Y_0-Y_1=\frac{\sqrt{X_0+2X_1}}{B(\frac{1}{3}\int Br\sqrt{X_0+2X_1}dr+C_2)},\label{Y0Y1Sol}
\end{equation}
and substituting (\ref{Y0Y1Sol}) in (\ref{Aest1}) we get
\begin{equation}
A=C_1\left (\int\frac{B r}{3}\sqrt{X_0+2X_1}dr+C_2\right ) \, , \label{Aest2}
\end{equation}
where again, $C_2$ is a constant of integration. 

Finally, the line element (\ref{SphericMetric}) can be rewritten as
\begin{equation}
ds^2=-C^2_1\left (\int \frac{1}{3}\sqrt{\frac{r^2 X}{1-\frac{r^2}{3}X}}dr+C_2\right )^2 dt^2+\frac{1}{1-\frac{r^2}{3}X}dr^2+
r^2(d\theta^2+sin^2\theta d\phi^2) \, , \label{Kmetric}
\end{equation}
which describes any Karmarkar static spherically symmetric anisotropic fluid distribution. Notice, that for this space-time we have defined 
\begin{equation}
X=X_0+2X_1=\frac{3}{r^3}\int X_0 r^2 dr\label{Xes} \, .
\end{equation}
Thus, all metrics will depend on a sole physical parameter: the energy density $X_0$ and in table \ref{TableMetricsX0} (see the Appendix at the end of the present work), we present the corresponding $X_0$ for several metrics which appeared in the recent literature.

To illustrate this strategy, let us assume the energy density as
\begin{equation}
    X_0=\frac{3+\frac{r^2}{R_s^2}}{(R^2_s+r^2)^2},
\end{equation}
then, from  equation (\ref{Xes}) we obtain that
\begin{equation}
    X=\frac{3}{r^2+R^2_s}
\end{equation}
and the line element (\ref{Kmetric}) can be written as follows
\begin{equation}
ds^2=-C^2_1\left(\frac{r^2}{2\sqrt{3}R^2_s}+C_2\right)^2 dt^2+ \left(1+\frac{r^2}{R^2_s}\right)dr^2+
r^2(d\theta^2+sin^2\theta d\phi^2)\label{Kemetric}
\end{equation}
which is the solution given in \cite{PandyaThomas2017}.

\subsection{$Z = 0$, the dynamic adiabatic scenario}
Much effort has been dedicated in developing static bounded Karmarkar models, but very little has been done for the dynamic case. In this section, we shall discuss the dynamic adiabatic state, $Z=0$, and explore the ``compatibility'' of the Karmarkar condition with other typical restrictions used in studying exact solutions in General Relativity.
\begin{itemize}
    \item \textbf{$X_1=0$, homogeneous energy density.} The uniform density spherical matter configuration is the standard entry point in all textbooks of General Relativity and Relativistic Astrophysics \cite{Weinberg1972,ShapiroTeukolsky1983,Schutz2009,MisnerThorneWheeler2017}. We have recently shown\cite{OspinoEtal2018} that, despite its simplicity and pedagogical interest, this widespread assumption is very restricted. Any dynamic homogeneous density profile satisfying the Karmarkar condition will lead to the Schwarzschild solution. It can be easily obtained from (\ref{Bianchi4}) assuming $X_0=X_0(t)$ and establishing regularity conditions at the origin we found $X_1=0$. Next, substituting this result into (\ref{KarmaKcondTetrad}), it leads to $Y_1=0$, i.e., conformally flat perfect fluid solution with homogeneous energy density: the Schwarzschild solution.
    \item \textbf{$Y_1=0$ vanishing complexity factor.} Recently, L. Herrera introduced a new concept of complexity for self-gravitating systems\cite{Herrera2018}. This concept includes the influences from energy density inhomogeneities and local anisotropy of the pressures on the active gravitational (Tolman) mass. Assuming the vanishing complexity condition, $Y_1=0$, in equation (\ref{KarmaKcondTetrad}) we obtain $X_1=0$, and because $Y_0\neq 0$, we re-obtain only the Schwarzschild solution.
    \item \textbf{$\mathcal{E}=0$, conformally flat case.} If $\mathcal{E}=0$, then   $X_1=Y_1$, and from (\ref{KarmaKcondTetrad}) we obtain $X_1=Y_1=0$, due to $X_0+Y_0\neq 0$ because the regularity at the origin.
    \item \textbf{$\Pi_1=0$, Pascalian isotropic fluids.} Karmarkar condition (\ref{KarmaKcondTetrad})  and the relation $X_1+Y_1=-8\pi \Pi_1$ lead to 
    \begin{equation}\label{KconX1}
     X_1=\frac{(Y_0-X_0-8\pi \Pi_1)-\sqrt{(Y_0-X_0-8\pi \Pi_1)^2-32\pi \Pi_1 X_0}}{2}
    \end{equation}
    and
    \begin{equation}\label{KconY1}
    Y_1=\frac{-(Y_0-X_0+8\pi \Pi_1)+\sqrt{(Y_0-X_0+8\pi \Pi_1)^2-32\pi \Pi_1 Y_0}}{2}
    \end{equation}
    Now, from  (\ref{KconX1}) and (\ref{KconY1}) it is clear that isotropy, $\Pi_1=0$ leads to $X_1=Y_1=0$, which is again the Schwarzschild solution.
    \item \textbf{$\sigma_1=\sigma_2$, shear-free and $a_1=0$ geodesic fluids.} Finally, considering shear-free and geodesic conditions in equations (\ref{ecR1}) y (\ref{ecR2}), we again obtain $Y_1=0$
\end{itemize}

\subsection{The dissipative case: $Z\neq 0$}
A recent paper \cite{NaiduGovenderMaharaj2018} develops a model of a radiating relativistic sphere that satisfies the Karmarkar condition. It is the first dynamic dissipative model obtained. 

For the present case, we shall consider a shear-free fluid, i.e.     
  \begin{equation}
  \sigma_1=\sigma_2=\sigma\,\,\, \Rightarrow \quad R=rB \, .
  \label{shfc}
  \end{equation}
From these two assumptions (Karmarkar and shear-free) we can find two new families of dissipative solutions.

If we assume that $X_1=0$, we can see immediately, from (\ref{ecR5}) and (\ref{ecR8}) that
\begin{equation}
J_1^\dag+\frac{1}{R^2}=0\,\,\,\Rightarrow \quad R=\frac{b(t)r}{1+a(t) b^2(t) r^2}
 \label{Rin}
\end{equation}
and from (\ref{ecR8}) that
\begin{equation}
\label{X0dissipative}
X_0 = 3\sigma^2 +12a(t) \, .
\end{equation}
Now, subtracting equation (\ref{Bianchi3}) from equation (\ref{Bianchi2}), we obtain
\begin{equation}
Y_1\sigma=Z^\dag+(2a_1-J_1)Z\label{Zeta1} \, .
\end{equation}
Next, by using the Kamarkar condition (\ref{KarmaKcondTetrad}) and taking into account (\ref{ecR3}),  equation (\ref{Zeta1}) can be written as:
 \begin{equation}
 \frac{X_0^\dag}{2X_0}=\frac{Z^\dag}{Z}+2a_1-J_1\,\,\, \Rightarrow \quad X_0=\frac{
 \tilde{x}_0(t) Z^2 A^4}{R^2}\label{X02}
 \end{equation}
where $\tilde{x}_0(t)$  is a constant of integration.

Thus, we can identify two possible cases:
\begin{enumerate}
  \item  For the case $a(t)=0$, from (\ref{ecR4}) we get $Z=a_1\sigma$ and by integrating (\ref{X02}) we find

 \begin{equation}\label{solant}
   A=C_2(t)\sqrt{r^2+C_1(t)},\quad R=b(t)r,\quad B=b(t)
 \end{equation}
which is the family of solutions shown in \cite{NaiduGovenderMaharaj2018}.

\item  For $a(t)=\frac{\tilde{a}}{b^2(t)}$, $x_0=\frac{2 b^2}{\dot{b}^2}$ and integrating (\ref{X02}) we find
  \begin{equation}
    A(t,r)=-\frac{\dot{b} \sqrt{C(-2+\tilde{a} C)+\tilde{a}r^4(-1+\tilde{a}^2 C^2)+2r^2(-1-\tilde{a}C+\tilde{a}^2 C^2)}}{\sqrt{2}(1+\tilde{a} r^2)}\label{solG}
  \end{equation}
  \noindent and
  \begin{equation}
  R(t,r)=\frac{b(t) r}{1+\tilde{a}r^2},\qquad B(t,r)=\frac{b(t) }{1+\tilde{a}r^2}\label{solRB}
  \end{equation}
\noindent with $C=\frac{4\tilde{c}(t)}{\dot{b(t)}}$, where $\tilde{c}(t)$ is a constant of integration.
\end{enumerate}
\noindent The matter variables for the solution (\ref{solG}) and (\ref{solRB}) are
\begin{equation}
    8\pi\rho=\frac{6}{b^2(t)}\left ( 2\tilde{a}+\frac{(1+\tilde{a} r^2)^2}{D1}\right )\qquad 8\pi \Pi=\frac{6r^2(1+\tilde{a}r^2)^2}{\dot{b}^2(t)D1^2} 
\end{equation}

\begin{equation}
    Z=\frac{2\sqrt{3}r(1+\tilde{a}r^2)}{b(t)\dot{b}(t) D1}\sqrt{2\tilde{a}+\frac{(1+\tilde{a} r^2)^2}{D1}}
\end{equation}
\begin{eqnarray}
8\pi P_r &=& \frac{\dot{C}(t)(1+\tilde{a}r^2)(1-\tilde{a}C(t)(1+\tilde{a}r^2))}{b(t)D1}-
\frac{\sqrt{2}(1+\tilde{a}r^2)C(t)\dot{b}(t)}{b^2(t)} \nonumber \\
        &-&\frac{4\tilde{a}}{b^2(t) D1}-\frac{10+8\tilde{a} r^2+6\tilde{a}^2r^2 +4\tilde{a}(-1+\tilde{a}^2r^4)}{b^2(t) D1^2}
\end{eqnarray}
\noindent with 
\begin{equation}
    D1=(r^2+C(t)+\tilde{a} C(t)r^2)(-2-\tilde{a}r^2+\tilde{a}C(t)(1+\tilde{a}r^2))
\end{equation}
\section{Final remarks}
\label{FinalRamarks}
It surprises the number of works published by slight variations in the metric functions. Then, after the integration of the Karamarkar condition (\ref{DiffKarmarkarStatic}), dozens of models (see Table \ref{TableMetricsX0}) are obtained with negligible or no discussion in their interrelations, remaining most of these efforts, in very descriptive stages.    

In this short article, we tried to explore some general consequences derived from the Karmarkar condition (\ref{KamarkarC}). By using a tetrad formalism in General Relativity and the orthogonal splitting of the Riemann tensor, we have presented it in terms of the structure scalars. Thus the new expression (\ref{KarmaKcondTetrad}) becomes an algebraic relation among the physical variables, and not a differential equation between the metric coefficients shown in (\ref{KarmarkarCondSchwComovDyna}) and (\ref{KarmarkarCondSchwCurvDyna}). Taking advantage of its simplicity, we have studied the static, dynamic adiabatic, and dynamic dissipative Karmarkar solutions. 

For the static case, we developed a method to obtain any spherically, static, anisotropic Karmarkar solution, parameterized by the energy density profile. We think it opens the possibility to explore new anisotropic matter configurations, starting from realistic isotropic nuclear equations of state.   

Much effort has been made on static bounded Karmarkar models, but very little considering the dynamic scenario. The simplicity of the scalar Karmarkar condition allows us to study the adiabatic and radiant cases efficiently.  Regarding the adiabatic dynamic matter configuration, we have shown that combining the Karmarkar condition with several other common simplifying assumptions, we inexorably obtain the homogeneous Schwarzschild solution. 

This raises a possible conjecture that for the spherical case, the Karmarkar dynamic adiabatic condition is incompatible with any other simplifying assumption. If they are combined, we necessarily obtain the homogeneous Schwarzschild solution.  This possible conjecture should be further, and carefully explored  in the future. 

Finally, for the dynamic dissipative case, we recovered a known previous solution \cite{NaiduGovenderMaharaj2018} and found a new shear-free Karmarkar radiating solution.

\section*{Acknowledgments}
J.O.  acknowledge financial support from Ministerio de Ciencia, Innovaci\'on y Universidades (grant PGC2018-096038-B-100) and Junta de Castilla y Le\'on  (grant SA083P17). J.O acknowledges hospitality of School of Physics of the Industrial University of Santander, Bucaramanga Colombia. L.A.N. gratefully acknowledge the financial support of the Vicerrector\'ia de Investigaci\'on y Extensi\'on de la Universidad Industrial de  Santander and the financial support provided by COLCIENCIAS under Grant No. 8863
\newpage
\section*{Appendix}
\begin{table}[h!]
    \centering
    \begin{tabular}{|l|c|} \hline \hline 
        $B^{2}$ &     $X_{0}$   \\ \hline \hline 
\cite{MauryaEtal2016B} 
         $1+{\frac{ \left( \alpha-\beta \right) {r}^{2}}{\beta {r}^{2}+1}}$ &
        ${\frac{ \left( \alpha-\beta \right)  \left( \alpha {r}^{2} +3 \right) }{ \left( \alpha {r}^{2}+1 \right)^{2}}}$  \\ \hline
\cite{SinghPant2016B} 
         $1+{\frac{\alpha {r}^{2}}{ \left( \beta {r}^{2}+1 \right)^{2}}}$ & 
         ${\frac{\alpha  \left( -{\beta}^{2}{r}^{4}+\alpha {r}^{2}+2 \beta  {r}^{2}+3 \right) }{ \left( {\beta}^{2}{r}^{4}+\alpha {r}^{2}+2 \beta  {r}^{2}+1 \right)^{2}}}$ \\ \hline        
\cite{SinghEtal2017} 
         $1 +{\frac {{\alpha}^{2}{r}^{2}}{{\beta}^{2}{r}^{4}+1}}$  & 
         ${\frac {{\alpha}^{2} \left( -{\beta}^{2}{r}^{4}+{\alpha}^{2}{r}^{2}+3 \right) }{ \left( {\beta}^{2}{r}^{4}+{\alpha}^{2}{r}^{2}+1 \right) ^{2}}}$ \\ \hline
\cite{Bhar2017} 
         $1 +\frac{{\alpha}^{2}{r}^{2}}{ \left( {\beta}^{2}{r}^{4}+1 \right)^{2}}$  &   
         ${\frac{{\alpha}^{2} \left(-5\,{\beta}^{4}{r}^{8}-2\,{\beta}^{2}{r}^{4}+{\alpha}^{2}{r}^{2}+3 \right) }{ \left( {\beta}^{4}{r}^{8}+2\,{\beta}^{2}{r}^{4}+{\alpha}^{2}{r}^{2}+1 \right)^{2}}}$ \\ \hline
\cite{BharSinghManna2017} 
         $1+{\frac {{\alpha}^{2}{r}^{2}}{ \left( \beta\,{r}^{2}+1 \right) ^{4}}}$  &   
         ${\frac {{\alpha}^{2} \left( -5\,{\beta}^{4}{r}^{8}-12\,{\beta}^{3}{r}^{6}-6\,{\beta}^{2}{r}^{4}+{\alpha}^{2}{r}^{2}+4\,\beta\,{r}^{2}+3 \right) }{ \left( {\beta}^{4}{r}^{8}+4\,{\beta}^{3}{r}^{6}+6\,{\beta}^{2}{r}^{4}+{\alpha}^{2}{r}^{2}+4\,\beta\,{r}^{2}+1 \right) ^{2}}}$ \\ \hline         
\cite{BharEtal2017} 
        $1+{\frac{{\alpha}^{2}{r}^{2}}{ \left( {\beta}^{2}{r}^{4}+ 1 \right)^{n}}}$ &
        ${\frac{ -4\,\left(  \left( -\frac{3}{4}+{\beta}^{2} \left( n-\frac{3}{4} \right) {r}^{4} \right)  \left( {\beta}^{2}{r}^{4}+1 \right)^{n}-1/4\,{\alpha}^{2}{r}^{2} \left( {\beta}^{2}{r}^{4}+1 \right)  \right) {\alpha}^{2}}{ \left( {\alpha}^{2}{r}^{2}+ \left( {\beta}^{2}{r}^{4}+1 \right)^{n} \right)^{2} \left( {\beta}^{2}{r}^{4}+1 \right) }}$ \\ \hline         
\cite{SinghPantGovender2017} 
       ${\frac{4 \left( \alpha {r}^{2}+1 \right)^{2}}{ \left(2-\alpha {r}^{2}\right)^{2}}}$  & 
        $ {\frac{3\alpha  \left( {\alpha}^{2}{r}^{4}+\alpha {r}^{2}+12\right) }{4 \left( \alpha {r}^{2}+1 \right)^{3}}}$\\ \hline
\cite{SinghBharPant2016} 
          $1+\alpha {r}^{2}+\beta {r}^{4}$ &    
          ${\frac{{\beta}^{2}{r}^{6}+2 \alpha \beta {r}^{4}+ \left( {\alpha}^{2}+5 \beta  \right) {r}^{2}+3 \alpha}{ \left( \beta {r}^{4}+\alpha {r}^{2}+1 \right)^{2}}}$ \\ \hline         
\cite{BharEtal2016} 
         $1 +64 \alpha {r}^{2} \left( \beta {r}^{2}+1 \right)^{2}$ & 
         ${\frac{ \left( 64 \alpha {\beta}^{3}{r}^{8}+192 \alpha {\beta}^{2}{r}^{6}+192 \alpha \beta {r}^{4}+64 \alpha {r}^{2}+7 \beta  {r}^{2}+3 \right)  \left( \beta {r}^{2}+1 \right) \alpha}{64\left( {\frac{1}{64}}+\alpha {r}^{2} \left( \beta {r}^{2}+1 \right)^{2} \right)^{2}} } $    \\ \hline
\cite{SinghPradhanPant2017}
         $1 +\alpha {r}^{2} \left( \beta {r}^{2}+1 \right)^{3}$  &   
         ${\frac{\alpha  \left( \beta {r}^{2}+1 \right)^{2} \left( 3+\alpha {\beta}^{4}{r}^{10}+4 \alpha {\beta}^{3}{r}^{8}+6 \alpha {\beta}^{2}{r}^{6}+4 \alpha \beta {r}^{4}+ \left( 9 \beta +\alpha \right) {r}^{2} \right) }{ \left( \alpha {\beta}^{3}{r}^{8}+3 \alpha {\beta}^{2}{r}^{6}+3 \alpha \beta {r}^{4}+\alpha {r}^{2}+1 \right)^{2}}}$ \\ \hline 
\cite{SinghPant2016}
         $1+ \alpha {r}^{2} \left( \beta {r}^{2} +1 \right)^{n}$ & 
         ${\frac{ -2\,\left(  \left( -\frac{3}{2}+\beta  \left( n-\frac{3}{2} \right) {r}^{2}\right)  \left( \beta {r}^{2}+1 \right)^{n}-\frac{1}{2} \alpha {r}^{2} \left( \beta {r}^{2}+1 \right)  \right) \alpha}{ \left( \alpha {r}^{2}+ \left( \beta {r}^{2}+1 \right)^{n} \right)^{2} \left( \beta {r}^{2}+1 \right) }}$   \\ \hline
 \cite{MauryaEtal2016C} 
         $1+ 4\,{n}^{2}\alpha {r}^{2} \left( \beta {r}^{2}+1 \right)^{n-2}$ & 
         ${\frac{\alpha {n}^{2} \left( \beta {r}^{2}+1 \right)  \left( \left( \beta {r}^{2}+1 \right)^{-1+2\,n}{n}^{2}\alpha {r}^{2}+\frac{1}{2}\, \left( \frac{3}{2}+\beta  \left( n-\frac{1}{2} \right) {r}^{2} \right)  \left( \beta {r}^{2}+1 \right)^{n} \right) }{ \left( {n}^{2}\alpha {r}^{2} \left( \beta {r}^{2}+1 \right)^{n}+1/4\, \left( \beta {r}^{2}+1 \right)^{2} \right)^{2}}}$  \\ \hline
\cite{SinghPantGovender2017B} 
        $\alpha r^{2}\sin^{2}\left(\beta r^{2} + \delta \right)$ &    
        ${\frac{\alpha {r}^{2} \left( \sin \left( \beta {r}^{2}+\delta \right)  \right)^{3}+4 \beta  \cos \left( \beta {r}^{2}+\delta \right) {r}^{2}+\sin \left( \beta {r}^{2}+\delta \right) }{\alpha {r}^{4} \left( \sin \left( \beta {r}^{2}+\delta \right)  \right)^{3}}}$ \\ \hline
\cite{SinghPantTroconis2017} 
       $1 +\alpha {r}^{2}  \cos^{2} \left( \beta {r}^{2} \right)$  & 
       ${\frac{{r}^{2}{\alpha}^{2} \left( \cos \left( \beta {r}^{2} \right) \right)^{4}-4 \alpha {r}^{2}\beta \sin \left( \beta {r}^{2} \right) \cos \left( \beta {r}^{2} \right) +3 \alpha  \left( \cos \left( \beta {r}^{2} \right)  \right)^{2}}{ \left( \alpha {r}^{2} \left( \cos \left( \beta {r}^{2} \right)  \right)^{2}+1 \right)^{2}}}$    \\ \hline 
\cite{Fuloria2017} 
         $1 +\frac{\alpha {r}^{2}}{ \cos^{4} \left( \beta {r}^{2}+\delta \right)}$  &    
         ${\frac{\alpha  \left( 8\, \left( \cos \left( \beta {r}^{2}+\delta \right)  \right)^{3}\sin \left( \beta {r}^{2}+\delta \right) \beta \,{r}^{2}+3\, \left( \cos \left( \beta {r}^{2}+\delta \right)  \right)^{4}+\alpha {r}^{2} \right) }{ \left(  \left( \cos \left( \beta {r}^{2}+\delta \right)  \right)^{4}+\alpha {r}^{2} \right)^{2}}} $ \\ \hline 
\cite{SinghMuradPant2017} 
         $1+\alpha {r}^{2}\tan \left( \beta {r}^{2}+\delta \right)$  &
         ${\frac{2 \alpha  \left( {r}^{2} \left( \beta+\alpha/2 \right)  \left( \tan \left( \beta {r}^{2}+\delta \right)  \right)^{2} +\frac{3}{2}\,\tan \left( \beta {r}^{2}+\delta \right) +\beta {r}^{2} \right) }{ \left( 1+\alpha {r}^{2}\tan \left( \beta {r}^{2}+\delta \right)  \right)^{2}}}$ \\ \hline        
\cite{MauryaRatanpalGovender2017} %
       $1+\delta\,{r}^{2} \left( \sinh \left( \alpha {r}^{2}+\beta \right) \right)^{2}$ &
       ${\frac{\delta\, \left(  \left( \cosh \left( \alpha {r}^{2}+\beta \right)  \right)^{2}\delta\,{r}^{2}-\delta\,{r}^{2}+3 \right) \left( \sinh \left( \alpha {r}^{2}+\beta \right)  \right)^{2}+4\,\delta\,\cosh \left( \alpha {r}^{2}+\beta \right) \alpha {r}^{2}\sinh \left( \alpha {r}^{2}+\beta \right) }{ \left(  \left( \cosh \left( \alpha {r}^{2}+\beta \right)  \right)^{2}\delta\,{r}^{2}-\delta\,{r}^{2}+1 \right)^{2}}} $ \\ \hline 
\cite{MauryaMaharaj2017} 
       ${\frac{1+2\,\delta\,{r}^{2}+\cosh \left( 2 \alpha {r}^{2}+2 \beta  \right) }{1+\cosh \left( 2 \alpha {r}^{2}+2 \beta  \right) }}$  &  
        ${\frac{2 \delta\, \left( -4\,\sinh \left( 2 \alpha {r}^{2}+2 \beta  \right) \alpha {r}^{2}+2\,\delta\,{r}^{2}+3\,\cosh \left( 2 \alpha {r}^{2}+2 \beta  \right) +3 \right) }{ \left( 1+2\,\delta\,{r}^{2}+\cosh \left( 2 \alpha {r}^{2}+2 \beta  \right)  \right)^{2}}}$ \\ \hline 
\cite{MauryaEtal2017} 
         $1+\alpha {r}^{2}\tanh \left( \beta {r}^{2}+\delta \right)$ &   
         ${\frac{-2 \alpha  \left( {r}^{2} \left( \beta-\alpha/2 \right)  \left( \tanh \left( \beta {r}^{2}+\delta \right)  \right)^{2}-\frac{3}{2}\,\tanh \left( \beta {r}^{2}+\delta \right) -\beta {r}^{2} \right) }{ \left( 1+\alpha {r}^{2}\tanh \left( \beta {r}^{2}+\delta \right)  \right)^{2}}}$ \\ \hline \hline
    \end{tabular}
    \caption{In this table, we present a non-exhaustive list of two dozen static metric functions that appeared in the literature between 2016 and 2017. These functions used to generate anisotropic equations of state via the Karmarkar differential condition (\ref{DiffKarmarkarStatic}), lead to a surprising number of mostly descriptive work done by small changes, and thus obtaining various models with negligible or no discussion in their interrelations.}
    \label{TableMetricsX0}
\end{table}

\newpage

\begin{thebibliography}{10}

\bibitem{OspinoHernandezNunez2017}
J.~{Ospino}, J.~L. {Hern{\'a}ndez-Pastora}, and L.~A. {N{\'u}{\~n}ez}.
\newblock An equivalent system of einstein equations.
\newblock {\em Journal of Physics Conference Series}, 831:012011, March 2017.

\bibitem{OspinoEtal2018}
J.~Ospino, J.L. Hern{\'a}ndez-Pastora, H.~Hern{\'a}ndez, and L.A.
  N{\'u}{\~n}ez.
\newblock Are there any models with homogeneous energy density?
\newblock {\em General Relativity and Gravitation}, 50(11):146, 2018.

\bibitem{karmarkar1948}
K.R. Karmarkar.
\newblock Gravitational metrics of spherical symmetry and class one.
\newblock {\em Proceedings of the Indian Academy of Sciences-Section A},
  27(1):56, 1948.

\bibitem{GuptaGupta1986}
Y.~K. Gupta and R.~S. Gupta.
\newblock Nonstatic analogues of kohler-chao solution of imbedding class one.
\newblock {\em General Relativity and Gravitation}, 18(6):641--648, Jun 1986.

\bibitem{Ivanov2017}
B.~V. {Ivanov}.
\newblock {Analytical study of anisotropic compact star models}.
\newblock {\em The European Physical Journal C.}, 77:738, 2017.

\bibitem{MauryaEtal2016B}
S.K. Maurya, Y.K. Gupta, T.T. Smitha, and F.~Rahaman.
\newblock A new exact anisotropic solution of embedding class one.
\newblock {\em The European Physical Journal A}, 52(7):191, 2016.

\bibitem{SinghPant2016B}
K.N. Singh and N.~Pant.
\newblock A new analytic solution representing anisotropic stellar objects in
  embedding class i.
\newblock {\em Astrophysics and Space Science}, page 176, 2016.

\bibitem{SinghEtal2017}
K.N. Singh, P.~Bhar, F.~Rahaman, N.~Pant, and M.~Rahaman.
\newblock Conformally non-flat spacetime representing dense compact objects.
\newblock {\em Modern Physics Letters A}, 32(18):1750093, 2017.

\bibitem{Bhar2017}
P.~Bhar.
\newblock Modelling a new class of anisotropic compact stars satisfying the
  karmakar's condition.
\newblock {\em The European Physical Journal Plus}, 132(6):274, 2017.

\bibitem{BharSinghManna2017}
P.~Bhar, K.N. Singh, and T.~Manna.
\newblock A new class of relativistic model of compact stars of embedding class
  i.
\newblock {\em International Journal of Modern Physics D}, 26(09):1750090,
  2017.

\bibitem{BharEtal2017}
P.~Bhar, K.N. Singh, N.~Sarkar, and F.~Rahaman.
\newblock A comparative study on generalized model of anisotropic compact star
  satisfying the karmarkar condition.
\newblock {\em The European Physical Journal C}, 77(9):596, 2017.

\bibitem{SinghPantGovender2017}
K.~N. Singh, N.~Pant, and M.~Govender.
\newblock Anisotropic compact stars in karmarkar spacetime.
\newblock {\em Chinese physics C}, 41(1):015103, 2017.

\bibitem{SinghBharPant2016}
K.N. Singh, P.~Bhar, and N.~Pant.
\newblock Solutions of the einstein's field equations with anisotropic pressure
  compatible with cold star model.
\newblock {\em Astrophysics and Space Science}, page 339, 2016.

\bibitem{BharEtal2016}
P.~Bhar, S.~K. Maurya, Y.~K. Gupta, and T.~Manna.
\newblock Modelling of anisotropic compact stars of embedding class one.
\newblock {\em The European Physical Journal A}, 52(10):312, 2016.

\bibitem{SinghPradhanPant2017}
K.N. Singh, N.~Pradhan, and N.~Pant.
\newblock New interior solution describing relativistic fluid sphere.
\newblock {\em Pramana}, 89(2):23, 2017.

\bibitem{SinghPant2016}
K.~N. Singh and N.~Pant.
\newblock A family of well-behaved karmarkar spacetimes describing interior of
  relativistic stars.
\newblock {\em The European Physical Journal C}, 76(10):524, 2016.

\bibitem{MauryaEtal2016C}
S.~K. Maurya, Y.~K. Gupta, S.~Ray, and D.~Deb.
\newblock Generalised model for anisotropic compact stars.
\newblock {\em The European Physical Journal C}, 76(12):693, 2016.

\bibitem{SinghPantGovender2017B}
Ksh.~N. Singh, N.~Pant, and M.~Govender.
\newblock Physical viability of fluid spheres satisfying the karmarkar
  condition.
\newblock {\em The European Physical Journal C}, 77(2):100, 2017.

\bibitem{SinghPantTroconis2017}
K.~N. Singh, N.~Pant, and O.~Troconis.
\newblock A new relativistic stellar model with anisotropic fluid in karmarkar
  space--time.
\newblock {\em Annals of Physics}, 377:256--267, 2017.

\bibitem{Fuloria2017}
P.~Fuloria.
\newblock Anisotropic compact star models in karmarkar space time continuum.
\newblock {\em Astrophysics and Space Science}, 362(12):217, 2017.

\bibitem{SinghMuradPant2017}
K.~N. Singh, M.~H. Murad, and N.~Pant.
\newblock A 4d spacetime embedded in a 5d pseudo-euclidean space describing
  interior of compact stars.
\newblock {\em The European Physical Journal A}, 53(2):21, 2017.

\bibitem{MauryaRatanpalGovender2017}
SK~Maurya, BS~Ratanpal, and M~Govender.
\newblock Anisotropic stars for spherically symmetric spacetimes satisfying the
  karmarkar condition.
\newblock {\em Annals of Physics}, 382:36--49, 2017.

\bibitem{MauryaMaharaj2017}
S.~K. Maurya and S.~D. Maharaj.
\newblock {Anisotropic fluid spheres of embedding class one using Karmarkar
  condition}.
\newblock {\em The European Physical Journal C}, 77(5):1--13, May 2017.

\bibitem{MauryaEtal2017}
S.K. Maurya, Y.K. Gupta, F.~Rahaman, M.~Rahaman, and A.~Banerjee.
\newblock Compact stars with specific mass function.
\newblock {\em Annals of Physics}, 385:532--545, 2017.

\bibitem{DebEtal2019}
D.~Deb, S.V. Ketov, S.K. Maurya, M.~Khlopov, P.~Moraes, and S.~Ray.
\newblock Exploring physical features of anisotropic strange stars beyond
  standard maximum mass limit in f(r,t) gravity.
\newblock {\em Monthly Notices of the Royal Astronomical Society},
  485(4):5652--5665, 2019.

\bibitem{AbbasNazar2019}
G.~{Abbas} and H.~{Nazar}.
\newblock {Stellar shear-free gravitational collapse with Karmarkar condition
  in f(R) gravity}.
\newblock {\em International Journal of Modern Physics A}, 34(33):1950220, Nov
  2019.

\bibitem{Wald2010}
R.~M. Wald.
\newblock {\em General relativity}.
\newblock University of Chicago press, 2010.

\bibitem{GarciaParrado2007}
A.~Garc{\'\i}a-Parrado G{\'o}mez-Lobo.
\newblock Dynamical laws of superenergy in general relativity.
\newblock {\em Classical and quantum gravity}, 25(1):015006, 2007.

\bibitem{HerreraEtal2009B}
L.~Herrera, J.~Ospino, A.~Di~Prisco, E.~Fuenmayor, and O.~Troconis.
\newblock Structure and evolution of self-gravitating objects and the
  orthogonal splitting of the riemann tensor.
\newblock {\em Physical Review D}, 79(6):064025, 2009.

\bibitem{HerreraEtal2014}
L~Herrera, A~Di~Prisco, J~Ib{\'a}{\~n}ez, and J~Ospino.
\newblock Dissipative collapse of axially symmetric, general relativistic
  sources: A general framework and some applications.
\newblock {\em Physical Review D}, 89(8):084034, 2014.

\bibitem{NaiduGovenderMaharaj2018}
N.~F. Naidu, M.~Govender, and S.~D. Maharaj.
\newblock Radiating star with a time-dependent karmarkar condition.
\newblock {\em The European Physical Journal C}, 78(1):48, 2018.

\bibitem{HerreraOspinoDiPrisco2008}
L.~Herrera, J.~Ospino, and A.~Di~Prisco.
\newblock {All static spherically symmetric anisotropic solutions of Einstein's
  equations}.
\newblock {\em Phys. Rev.}, D77:027502, 2008.

\bibitem{PandyaThomas2017}
D.M. Pandya and V.O. Thomas.
\newblock Models of compact stars on paraboloidal spacetime satisfying
  karmarkar condition.
\newblock {\em arXiv preprint arXiv:1708.06220}, 2017.

\bibitem{Weinberg1972}
S.~Weinberg.
\newblock {\em Gravitation and Cosmology: Principles and Applications of the
  General Theory of Relativity}.
\newblock Wiley, 1972.

\bibitem{ShapiroTeukolsky1983}
S.~L. Shapiro and S.~A. Teukolsky.
\newblock {\em {The Physics of Compact Objects}}.
\newblock Wiley, New York, 1983.

\bibitem{Schutz2009}
B.~F. Schutz.
\newblock {\em {A First Course in General Relativity}}.
\newblock Cambridge University Press, 2009.

\bibitem{MisnerThorneWheeler2017}
C.~W. Misner, K.~S. Thorne, and J.~A. Wheeler.
\newblock {\em Gravitation}.
\newblock Princeton University Press, 2017.

\bibitem{Herrera2018}
L.~Herrera.
\newblock New definition of complexity for self-gravitating fluid
  distributions: The spherically symmetric, static case.
\newblock {\em Physical Review D}, 97(4):044010, 2018.

\end{thebibliography}

\end{document}